\documentclass[conference,letterpaper]{IEEEtran}

\usepackage[utf8]{inputenc} 
\usepackage[T1]{fontenc}
\usepackage{url}
\usepackage{ifthen}
\usepackage{ifpdf}
\usepackage{cite}
\usepackage[cmex10]{amsmath} 
\usepackage{lipsum}
\usepackage{booktabs}
\usepackage{amssymb}
\usepackage{bm,bbm}
\usepackage[inline,shortlabels]{enumitem}
\usepackage{threeparttable}
\usepackage[ruled,norelsize]{algorithm2e}
\usepackage{xcolor}
\usepackage{subfigure}
\usepackage{pgf,tikz}
\DeclareGraphicsExtensions{.pdf,.jpeg,.png}

\makeatletter
\newcommand{\removelatexerror}{\let\@latex@error\@gobble}
\makeatother

\newtheorem{definition}{Definition}[section]
\newtheorem{proposition}{Proposition}[section]
\newtheorem{corollary}{Corollary}[section]
\newtheorem{theorem}{Theorem}[section]

\begin{document}

\title{GRAND Massive Parallel Decoding Framework for Low Latency in Beyond 5G}

\author{%
 \IEEEauthorblockN{Danilo Gligoroski, Sahana Sridhar, Katina Kralevska}
 \IEEEauthorblockA{Department of Information Security and Communication Technology\\
                   Norwegian University of Science and Technology (NTNU) -
                   Trondheim, Norway\\
                   Email: \{danilog, sahana.sridhar, katinak\}@ntnu.no}
}



\maketitle


\begin{abstract}
   We propose a massive parallel decoding GRAND framework. The framework introduces two novelties: 1. A likelihood function for $M$-QAM demodulated signals that effectively reduces the symbol error pattern space from $\mathcal{O}(5^{N/\log_2 M})$ down to $\mathcal{O}(4^{N/\log_2 M})$; and 2. A massively parallel matrix-vector multiplication for matrices of size $K\times N$ ($K \leq N$) that performs the multiplication in just $\mathcal{O}(\log_2 N)$ steps. We then apply the proposed GRAND approach to codes and operational modulation techniques used in the current 5G NR standard. Our framework is applicable not just to short codewords but to the full range of codewords from 32 bits up to 1024 bits used in the control channels of 5G NR. We also present simulation results with parity-check matrices of Polar codes with rate $R=1/2$ obtained from the 5G NR universal reliability sequence.

\end{abstract}

\section{Introduction}

The next generation of cellular systems, beyond 5G and 6G, have been designed to meet the ever-increasing demand for new applications with various requirements. One of the approaches to fulfil some of the key 5G requirements, i.e. over-the-air latency ($< 1ms$), end-to-end latency ($< 5ms$) and peak throughput ($\approx 20$ Gbps), is to reduce the \emph{structural} and \emph{processing} delays of channel coding.

In 2009, Arikan~\cite{arikan2009channel} proposed Polar codes, showing that they are provably capacity-achieving in binary discrete memoryless channels (B-DMCs) for large codeword lengths ($N \to \infty$). The channel polarization phenomenon transforms the bit-channels to either completely noise-free or extremely noisy channels. The information bits are sent over the $K$ most reliable \emph{``good'' information} channels, while the remaining $N-K$ \emph{``bad'' frozen} channels are set to a fixed value or zero. Polar codes have been adopted by 3GPP for the control channel in 5G New Radio (NR). 

Additionally, low complexity encoding and decoding algorithms, i.e. successive cancellation decoding (SCD), were also proposed in~\cite{arikan2009channel}. However, error correction performance degrades for shorter code lengths and induces delays due to the inherent sequential nature of SCD. Later, the SC list (SCL) decoding~\cite{tal2015list} was developed, where a list is maintained during decoding, and the most reliable codeword among $L$ candidate codewords is selected in the end. This improves the block error rate (BLER) but induces additional latency. Moreover, a cyclic redundancy check (CRC) appended to the codeword helps to select the most reliable codeword and to further improve the BLER performance~\cite{niu2012crc}. Improvements for the throughput, latency and complexity of SCD have been proposed, such as the simplified successive cancellation decoding (SSC)~\cite{alamdar2011simplified}, its faster variants (F-SSC)~\cite{hanif2017fast}, specific decoders for the constituent codes of the decoding tree~\cite{alamdar2011simplified} and its extensions to the SCL decoder~\cite{hashemi2017fast}.

While these algorithms contribute to throughput and latency improvements, they still fall behind when it comes to practicability due to either increased complexity or limitations in parallelization. Therefore, there is a need to look for new solutions that can provide maximal parallelization with low complexity for a range of code rates and various block lengths. 

The recently introduced Guessing Random Additive Noise Decoding (GRAND) algorithm~\cite{duffy2019capacity} has attracted the attention of the coding theory community due to its efficient error correction performance -- for short block codes and high code rates -- and its highly parallelizable properties. GRAND is a noise guessing algorithm that initially generates the most likely test error patterns (TEPs), applies them to the hard demodulated received signal and checks whether the resulting string is a member of the codebook. It makes use of ordered statistics of noise for decoding, and the codebook is only used for the look-up of the potential decoded codeword in the codebook. The noise statistics may be obtained by arbitrary means and are not dependent on examining the output from the decoder.

\begin{table*}[!htbp]
\renewcommand{\arraystretch}{1.3}
\centering
\begin{threeparttable}[H]
\caption{Comparison with proposed GRAND algorithms over higher-order modulation techniques as used in 5G NR.}
\label{tab:contrib}
\begin{tabular}{|l||l|l|l|l|l|}
\hline
 & \parbox{1.5cm}{Channel} & \parbox{3.0cm}{Modulation} & Block size $N$ & \parbox{2.5cm}{\vspace{0.1cm}Size of symbol error pattern space\vspace{0.1cm}}  & \parbox{2.5cm}{Parallel execution in \# steps}\\ \hline \hline
\parbox{3.0cm}{Fading-GRAND~\cite{hadi2022fading} } & \begin{tabular}[l]{@{}l@{}}Rayleigh\\ Rician\end{tabular} & \parbox{3.0cm}{QPSK, 16-QAM,\\ 64-QAM} & $127, 128$ & NA & NA  \\ \hline
\parbox{3.0cm}{GRAND-for-NNE~\cite{an2022burst}} & \parbox{1.5cm}{\vspace{0.1cm} Memoryless channels \vspace{0.1cm}} & \parbox{3.0cm}{16-QAM, 64-QAM,\\ 256-QAM} & $127, 128$ & $\mathcal{O}(5^{N/\log_2M})$ & NA \\ \hline
\parbox{3.0cm}{\textbf{\textit{This Work}}} & AWGN & \parbox{3.0cm}{\vspace{0.1cm} $M$-QAM, $M \in \{4,$ $16$, $64$, $256$, $1024$, $4096\}$ \vspace{0.1cm} } & \parbox{3.0cm}{\vspace{0.1cm} All 5G control channel block lengths: 32, 64, 128, 256, 512, 1024 \vspace{0.1cm}} & $\mathcal{O}(4^{N/\log_2M})$ & \parbox{2.5cm}{$2n + 2S + 4$ \ \  \tnote{i} }  \\ \hline
\end{tabular}
\begin{tablenotes}
    \item [i] $n = \log_2 N$, $S$ is the cut-off parameter, and in our experiments $S = 8$
\end{tablenotes}
\end{threeparttable}
\end{table*}

Several variants of GRAND have been devised based on the method and order of generation of the test patterns. GRAND with ABandonment (GRAND-AB)~\cite{duffy2019capacity} is a hard decision input variant that generates TEPs in ascending Hamming weight order, up to the weight AB. Two soft-input variants that leverage soft information are Ordered Reliability Bits GRAND (ORBGRAND)~\cite{duffy2022ordered} -- which uses \emph{logistic weight} to order the TEPs; and Soft GRAND (SGRAND)~\cite{duffy2019guessing} -- which maintains a dynamic candidate error set during TEP generation, resulting in improved decoding performance compared to that of the hard-input GRANDAB. Both hard-detection (GRAND-SOS)~\cite{an2021crc} and soft detection (ORBGRAND) variants have been used to decode short, high-rate block codes and also for error correction of CRC-coded data. There is also List-GRAND algorithm~\cite{abbas2022list}, which generates a list of $L$ estimated codewords to select the most likely candidate. Some works~\cite{abbas2022fading,hadi2022fading,chatz2023fading} have applied GRAND to fading channels.
In comparison to other code-agnostic channel code decoders, such as brute-force ML decoding and ordered statistic decoding, GRAND and its variants offer a relatively low-complexity decoding solution for short-length and high-rate channel codes. 


\subsection{Contributions}

The contributions of this paper are listed as follows:
\begin{enumerate}
    \item We focus on high-order modulation techniques used in 5G NR, such as $M$-QAM for \(M \in \{4, 16, 64, 256, 1024, 4096\}\) and in the AWGN channel.
    \item Unlike published GRAND approaches that aim to achieve SNR gains and improved decoding success probability (but also slow down the decoding process), we focus exclusively on the maximum parallelizable approach.  
    \item We apply the proposed approach to all block lengths used in 5G control channels, i.e. for $N \in \{32, 64, 128, 256, 512, 1024\}$.
    \item We introduce a likelihood function for $M$-QAM demodulated signals that reduces the size of symbol error pattern space from $\mathcal{O}(5^{N/\log_2M})$ to $\mathcal{O}(4^{N/\log_2M})$.
    \item We describe a novel massively parallel matrix-vector multiplication. In the context of Polar codes defined for 5G NR, the matrix is the parity-check matrix $\mathbf{H}_{K\times N}$ for the corresponding Polar code. Our parallel algorithm performs the vector-matrix multiplication in $\log_2 N $ steps. Simulation results are presented for the code rate $R=1/2$, i.e. for $K = N/2$.
\end{enumerate}

Table \ref{tab:contrib} summarizes the comparison of our work with two state-of-the-art GRAND approaches for higher-order QAM techniques.

The rest of the paper is organized as follows: Section II presents background and preliminaries that form the basis for the components of the proposed massive parallel decoding framework. Section III explains the proposed massive parallel decoder. Section IV presents performance evaluation. Section V concludes the paper. 

\section{Preliminaries}

In this section, we present some background and preliminaries that form the basis for the components of the proposed massive parallel decoding framework.  


\begin{definition}[Polar codes~\cite{arikan2009channel}]
    A $(N,K)$ Polar code, with block length $N = 2^n$ and code dimension $K$, is specified by a generator matrix $\mathbf{G}_N$, which is the $n$\textsuperscript{th} Kronecker product of the binary kernel $\mathbf{F}$, 
    \[\mathbf{G}_N = \mathbf{F}^{\otimes n} \mbox{ where } \mathbf{F} = \begin{pmatrix}
        1 & 0\\
        1 & 1
    \end{pmatrix}.\]
    The input vector $\mathbf{u} = (u_0, u_1, \dots, u_{N-1})$, consisting of \emph{information} $\mathcal{I}$ and \emph{frozen} $\mathcal{F}$ bits 
    is encoded as $\mathbf{c} = \mathbf{u} \cdot \mathbf{G}_N$, to produce the codeword $\mathbf{c} = (c_0, c_1, \dots, c_{N-1})$. 

    Given the $K\times N$ matrix $\mathbf{G}_\mathcal{I}$ that is obtained from $\mathbf{G}_N$ by keeping the rows with indices in $\mathcal{I}$, the parity check matrix $\mathbf{H}$ is a $(N-K)\times N$ matrix whose columns generate the Null space for $\mathbf{G}_\mathcal{I}$ 
    \[ \mathbf{H} = \mathsf{NullSpace}(\mathbf{G}_\mathcal{I}) \mbox{, i.e. } \mathbf{G}_\mathcal{I} \mathbf{H}^T = 0. \]
\end{definition}

%

In determining the sets $\mathcal{I}$ and $\mathcal{F}$, Arikan in \cite{arikan2009channel} used Bhattacharyya parameters as a measure for the channel reliability and showed that for Binary Erasure Channels and Binary Symmetric Channels, finding the optimal sets of $\mathcal{I}$ and $\mathcal{F}$ depends on the channel erasure, i.e. error probability. 

When adopting Polar codes for the control channels in 5G NR \cite{etsi2020138}, 3GPP aimed to construct Polar codes that will cover a broader range of code lengths and channels with different error probabilities. For block lengths of $N = 2^n$ where $n\in \{5, 6, 7, 8, 9, 10\}$, 3GPP defined a universal reliability sequence $Q_{0}^{N_{\text{max}}-1}$ where $N_{\text{max}}=1024$, sorted in ascending order of the reliability of the bits (given in Table 5.3.1.2-1 of \cite{etsi2020138}). 



Next, in line with our used notation so far, we give the following definition of square Quadrature Amplitude Modulation of size $M$ ($M$-QAM) (adapted from Section 3.2-3 of~\cite{john2008digital}):
\begin{definition}
    Let $M = 2^{2m}$ be $M$ symbols represented as bit sequences of $2m$ bits. A square $M$-QAM signal carried by two quadrature carriers: cosine carrier $\cos{\omega_c t}$ and sine carrier $\sin{\omega_c t}$ (with angular frequency $\omega_c = 2\pi f_c$ and  carrier frequency $f_c$) is represented as:
    \begin{multline}
    \label{Eq:QAM-signal}
        s_i(t) = A_{ic} \cos{\omega_c t} - A_{is} \sin{\omega_c t}, \\ 0\leq t \leq T_s, i = 1, \ldots, M
    \end{multline}
\end{definition}
where $T_s$ is the time interval for sending one symbol. The quadrature amplitudes $A_{ic}$ for the cosine carrier and $A_{is}$ for the sine carrier, take values from the set $\{\pm d, \pm 3d, \pm 5d, \ldots \}$, where $2d$ is the minimum Euclidean distance of a set $\mathit{Const}_d$ of $M$ constellation points $\mathit{Const}_d = \{(A_{1c}, A_{1s}), (A_{2c}, A_{2s}), \ldots, (A_{Mc}, A_{Ms}) \} $, where $d$ is given by the expression
\begin{equation}
\label{Eq:minEuclidDistance}
    d = \sqrt{\frac{3m E_b}{M-1}}
\end{equation}
and $E_b$ is the average energy per bit. Since every point $(A_{ic}, A_{is}) \in \mathit{Const}_d$ has the common scaling factor $d$, in the literature, it is more convenient to represent the constellation as the set  $\mathit{Const} = \{(A_{1c}, A_{1s}), (A_{2c}, A_{2s}), \ldots, (A_{Mc}, A_{Ms}) \} $ where $A_{ic}$ and $A_{is}$ are odd integers, i.e. $A_{ic}, A_{is} \in \{\pm 1, \pm 3, \pm 5, \ldots \}$. 

\begin{figure}
    \centering
    \includegraphics[width=0.5\textwidth]{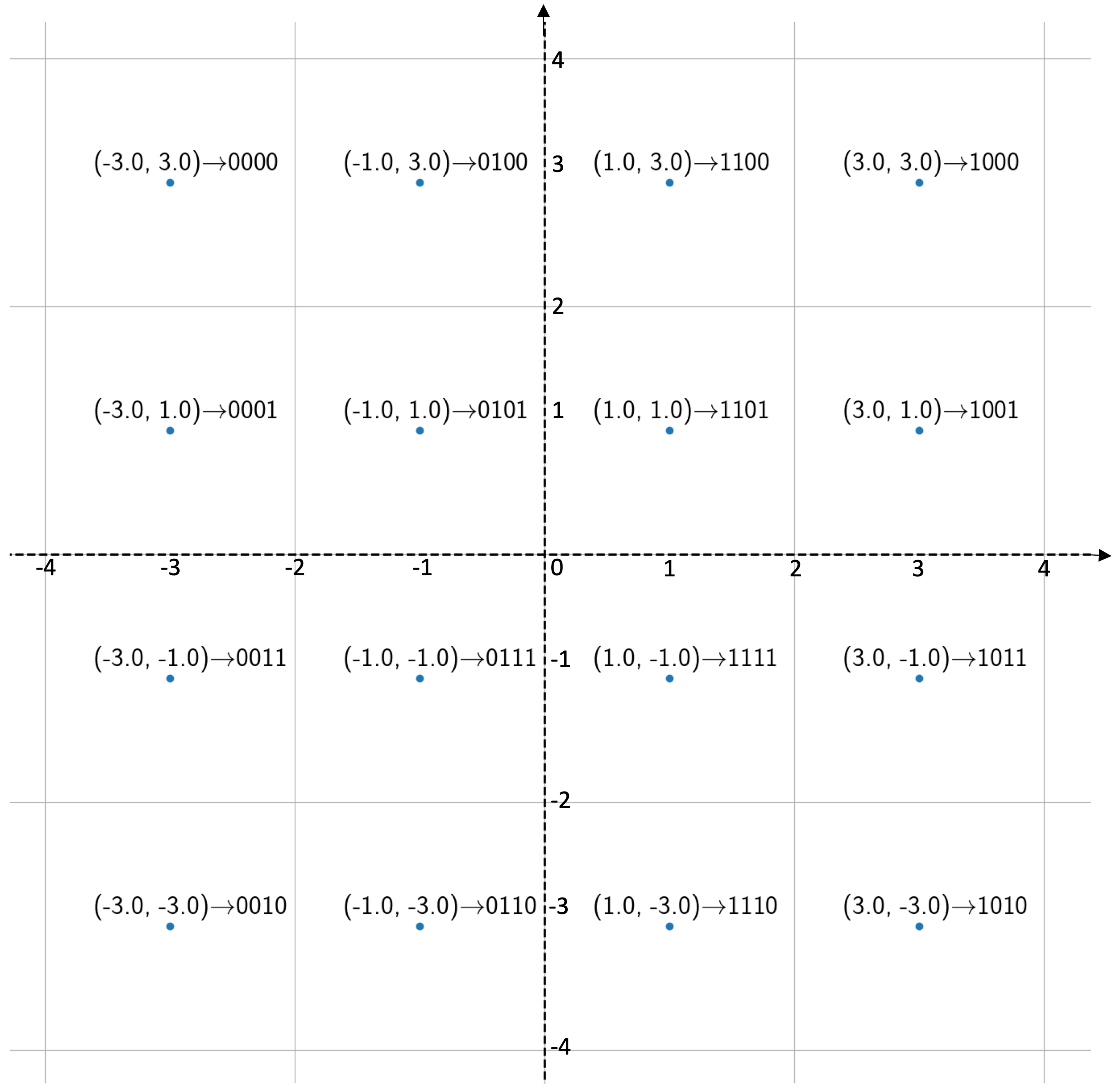}
    \caption{A 16-QAM constellation with associated Gray code for 4-bit sequences.}
    \label{fig:16-QAM-Constellation}
\end{figure}

In Figure \ref{fig:16-QAM-Constellation}, we present a 16-QAM constellation with $A_{ic}, A_{is} \in \{\pm 1, \pm 3\}$ together with the corresponding 4-bit Gray code sequences. 

Next, we briefly introduce some notions specific to GRAND. We adapt the modelling of GRAND for higher-order QAM modulation given in \cite{an2022burst} to be consistent with the notation. Consider a $M$-QAM modulation, and an alphabet $\mathbb{A}$ that has $M$ elements, i.e. $|\mathbb{A}| = M$. We want to send $L$ symbol codeword $X^L$ over a channel with arbitrary rate $R$, where the codebook consists of $M^{R L}$ strings from $\mathbb{A}^L$. Under the assumption that every codeword symbol is affected independently by the channel noise, we have a channel noise $Noise^L$, which also takes values from $\mathbb{A}^L$. $Noise^L$ alters $X^L$ symbol-wise between transmission and reception. Upon reception, we have a resulting sequence $Y^L = X^L \odot Noise^L$ where $\odot$ represents the \emph{alteration} operation that replaces the original symbol in $X^L$ with another symbol from $\mathbb{A}$.

GRAND approach is an attempt to determine $X^L$ from $Y^L$ by identifying $Noise^L$ through producing the most likely test error patterns $Z^L$ and applying an inverse operation $\circledcirc$ (to the operation $\odot$) to the received signal and querying if the result, $Y^L \circledcirc Z^L$, is in the codebook $\mathcal{C}$. 

In \cite{duffy2019guessing}, the authors proved that if transmitted codewords are equally likely and the test error patterns $Z^L$ are queried in order from most likely to least likely based on the channel’s statistics, the first instance where a codebook element is found is a maximum likelihood (ML) decoding. Additionally, they proved that if the search is abandoned after a certain number of queries, although it is not an ML decoding, it is also a capacity-achieving decoding technique for an appropriate choice of abandonment threshold.

For determining the $Noise^L$ with $M$-QAM modulation, the authors of \cite{an2022burst} worked in the "Nearest Neighbor Error" (NNE) channel and assumed that hard detection errors only occur with immediate neighbours. Due to the nature of $M$-QAM constellations in the complex plane, the four error directions are "north", "south", "east", and "west" (except when the hard decoded symbol is on the edge or in the corner of the constellation). This results in the size of symbol error pattern space being $\mathcal{O}(5^{N/\log_2M})$.


\section{The proposed massive parallel decoder} 
\label{Sec:OurProposal}

The codes used in the experiments and simulations in this work have a rate of $R = 1/2$, block lengths of $N = 2^n$ bits, and for those parameters, the parity check matrices are obtained from the universal reliability sequence $Q_{0}^{N_{\text{max}}-1}$. We do not use puncturing and rate matching. Additionally, since the GRAND-like decoding approaches are agnostic to the specifics of the code, we do not use the CRC and PC bits.

First, we introduce the likelihood function $\mathcal{L}(r, s)$.
\begin{definition}
    \label{def:LikelihoodFunctionAndNNEs}
    Let $r = (a, b)$ be a received signal for which the $M$-QAM demodulator returns the hard-decoded constellation point with coordinates $s = (A_{ic}, A_{is}) $ in the complex plane. Let 
    \begin{align}
        d_1 & = \left\{
                \begin{array}{rl}
                  1 - (A_{ic} - a) & \text{if } (A_{ic} - a) > 0,\\
                  1 + (A_{ic} - a) & \text{if } (A_{ic} - a) < 0.
                \end{array} \right. \\
        d_2 & = \left\{
                \begin{array}{rl}
                  1 - (A_{is} - b) & \text{if } (A_{is} - b) > 0,\\
                  1 + (A_{is} - b) & \text{if } (A_{is} - b) < 0.
                \end{array} \right.
    \end{align}
    The likelihood function $\mathcal{L}(r, s)$ is defined as
    \begin{equation}
        \mathcal{L}(r, s) = \sqrt{d_1^2 + d_2^2} \ \ .
    \end{equation}
\end{definition}

\begin{figure}
    \centering
    \includegraphics[width=0.5\textwidth]{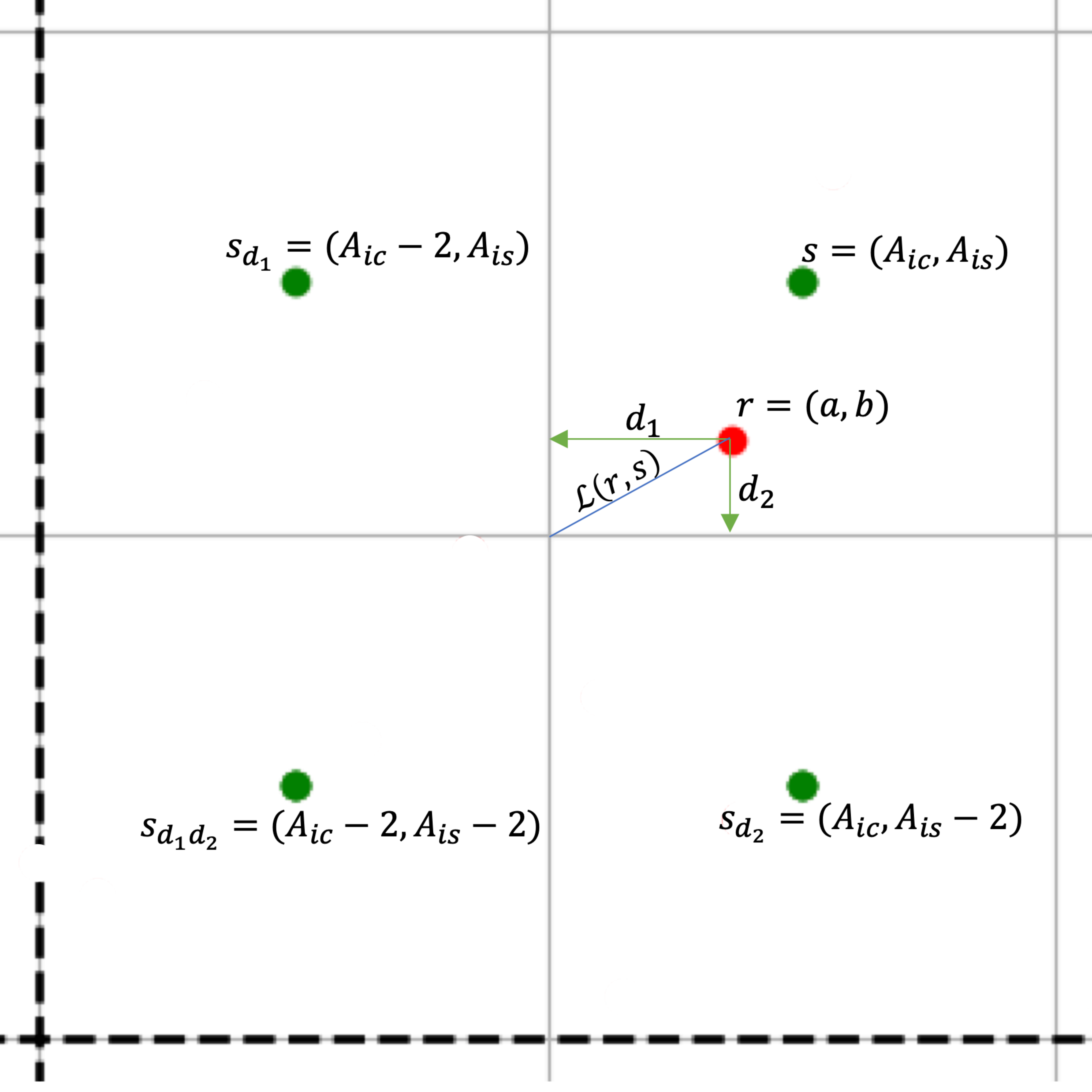}
    \caption{Likelihood function $\mathcal{L}(r, s)$ and Near Neighbours Errors in our approach.}
    \label{fig:LikelihoodFunction}
\end{figure}

In Figure \ref{fig:LikelihoodFunction}, we give an example of computing the likelihood function $\mathcal{L}(r, s)$ from a received signal $r = (a, b)$ and its demodulated hard-decoded value $s = (A_{ic}, A_{is}) $.
\begin{definition}
    \label{def:OurNNEs}
    Let $d_1$ and $d_2$ be calculated as in Definition \ref{def:LikelihoodFunctionAndNNEs}. Then, up to three Near Neighbours Errors constellation points - "horizontal" $s_{d_1}$, "vertical" $s_{d_2}$ and "diagonal" $s_{{d_1}{d_2}}$ are defined as:
    \begin{align}
        \left\{
            \begin{array}{rl}
            s_{d_1} & = s - (2\ \mathsf{Sign}(A_{ic} - a), 0),\\
            s_{d_2} & = s - (0, 2\ \mathsf{Sign}(A_{is} - b)),\\
            s_{{d_1}{d_2}} & = s - (2\ \mathsf{Sign}(A_{ic} - a), 2\ \mathsf{Sign}(A_{is} - b))
        \end{array} \right.
    \end{align}
   if the calculated coordinates of $\{s_{d_1}, s_{d_2}, s_{{d_1}{d_2}}\}$ belong to the constellation $\mathit{Const}$.
\end{definition}

Similar to reference \cite{an2022burst}, we can have situations when the hard-decoded constellation point is at the edge or in the corner of the constellation. However, instead of having up to five potential points for most likely test error patterns (the hard-decoded point plus "north", "south", "east", and "west" points), in our case, we have four potential points for most likely test error patterns: the hard-decoded point plus "horizontal", "vertical" and "diagonal" points. This discussion is summarized with the following.
\begin{corollary}
    The size of symbol error pattern space in our approach is $\mathcal{O}(4^{N/\log_2M})$.
\end{corollary}

\begin{figure}
    \centering
    \includegraphics[width=0.5\textwidth]{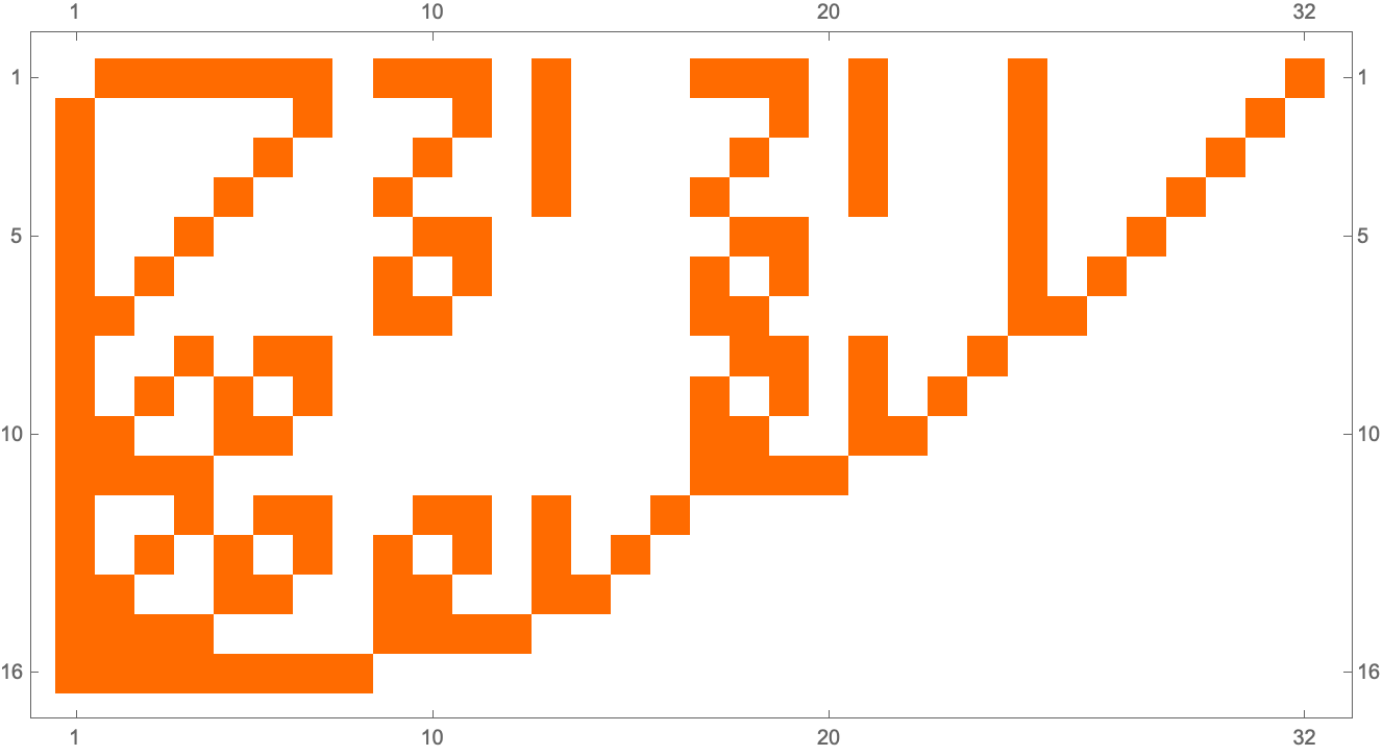}
    \caption{The parity check matrix $\mathbf{H}_{32}$ of dimensions $16 \times 32$. Every coloured square represents 1, and every blank square represents 0.}
    \label{fig:H32}
\end{figure}

Let the block length be $N = 2^n$ bits where $n\in \{5, 6, 7, 8, 9, 10\}$, and let the rate be $R=1/2$. Using the 3GPP universal reliability sequence $Q_{0}^{N_{\text{max}}-1}$, let the matrices $\mathbf{H}_N$ of size $(\frac{N}{2})\times N$ be the corresponding parity check matrices. The parity check matrix $\mathbf{H}_{32}$ is graphically presented in Figure \ref{fig:H32}.

Our first observation about the matrices $\mathbf{H}_N$ is that their first row has the highest Hamming weight. Second, as $n$ increases, the sparsity of the matrices increases. These two observations guide us in defining the following matrix-vector multiplication.

For defining the massive parallel matrix-vector multiplication, we will use the following notation: Let $\mathbf{H}_N = [h_{ij}]$ with elements $h_{ij}$ be a parity check matrix of a Polar code with rate $R=1/2$ defined for a block length $N$ with the universal reliability sequence $Q_{0}^{N_{\text{max}}-1}$, let the Hamming weight of the $i$-th row of $\mathbf{H}_N$ be $W_\mathbf{H}(i)$ and let the total Hamming weight of the matrix be $W_\mathbf{H}$. Let $\mathbf{H}_{\mathcal{I}(i)}$ be the index set for the $i$-th row, i.e. if $j \in \mathbf{H}_{\mathcal{I}(i)}$ then $h_{ij}=1$.

\begin{table}[h!]
	\centering
	\begin{tabular}{|c|}
		\hline
		\parbox{8.5cm}{\vspace{0.1cm} \textbf{Algorithm 1:} Massive Parallel Matrix-Vector multiplication\\
                     $\mathsf{pMult}(\mathbf{H}, \mathbf{c}^T)$ \vspace{0.1cm}}\\
		\hline
		\parbox{8.5cm}{\vspace{0.1cm}{\bf Input:} Matrix $\mathbf{H}$ of dimensions $K\times N$ and a row vector $\mathbf{c}$ of dimension $N$.\\
			{\bf Output:} A vector column $\mathbf{r} = \mathbf{H} \mathbf{c}^T$ of dimension $K$.
			\vspace{0.1cm}} \\
		\hline
		\parbox{8.5cm}{
			\begin{description}
				\item[Step 1:] Parallel execution for each $i$ and each $j$: For each $i \in 1, \ldots \frac{N}{2}$ and         each $j \in \mathbf{H}_{\mathcal{I}(i)}$, $r_{ij} \leftarrow c_j h_{ij} $ \\
				Rename variables (this is not executable action - just step convention): $r_{ij'} \leftarrow r_{ij}$ for        $j' \in 1,\ldots,W_\mathbf{H}(i)$ and\\ 
                    $\mathbf{H}_{\mathcal{I}}(i) \leftarrow \{1,\ldots,W_\mathbf{H}(i)\}$
				\item[While $|\mathbf{H}_{\mathcal{I}}(i)|>1$ for any $i$]
				
				        \item \begin{description}
					    \item[\ \ Step 2:] Parallel execution for each $i$ where $|\mathbf{H}_{\mathcal{I}}(i)|>1$ and each        odd $j \in \mathbf{H}_{\mathcal{I}}(i)$:  $r_{ij} \leftarrow r_{ij} + r_{i(j+1)} $ \\
					    Rename variables (this is not executable action - just step convention): $r_{ij'} \leftarrow                r_{ij}$ for $j' \in 1,\ldots, \lfloor\frac{W_\mathbf{H}(i)}{2}\rfloor $ and\\
                            $\mathbf{H}_{\mathcal{I}}(i) \leftarrow \{1,\ldots,\lfloor\frac{W_\mathbf{H}(i)}{2}\rfloor\}$
				\end{description}
				
				\item[End while] 
				\item[Return] $\mathbf{r} = (r_{11}, r_{21},\ldots, r_{K1} )$
			\end{description}
		}\\
		\hline
	\end{tabular}
	\label{Alg:pMult}
\end{table}

\begin{theorem}
    \label{Thm:ParallelMatrixVectorMultiplicationTime}
    Let a candidate codeword be the (row) vector $\mathbf{c} = (c_0, c_1, \dots, c_{N-1})$. Then the matrix-vector multiplication \(\mathbf{H} \mathbf{c}^T\) with Algorithm 1 takes 
    \begin{equation}
        \label{Eq:pMultComplexity}
        1 + \max_{i \in 1, \ldots K} \lceil \log_2( W_\mathbf{H}(i)) \rceil 
    \end{equation} parallel steps. The total number of logical $\mathsf{AND}$ gates (for all bit multiplications) is $W_\mathbf{H}$, and the total number of logical $\mathsf{XOR}$ gates (for all bit additions) is 
    \begin{equation}
        \sum_{i = 1}^{K} \lceil \log_2( W_\mathbf{H}(i)) \rceil.    
    \end{equation}
\end{theorem}

In Table \ref{Table:SummaryHN} we summarize the findings from previous analysis for all $\mathbf{H}_N$.

\begin{table}[htbp]
\caption{Summary table.}
\label{Table:SummaryHN}
\resizebox{\columnwidth}{!}{%
\begin{tabular}{@{}|l|r|r|r|r|r|r|@{}}
\toprule
$n$                                  & 5  & 6   & 7   & 8   & 9    & 10   \\ \midrule
$N = 2^n$                            & 32 & 64  & 128 & 256 & 512  & 1024 \\ \midrule
$W_\mathbf{H} = \mathsf{AND}$ gates       & 136     & 322     & 984     & 2890   & 8322   & 24828  \\ \midrule
Sparsity $\frac{W_\mathbf{H}}{K\times N}$ & 26.56\% & 15.72\% & 12.01\% & 8.82\% & 6.35\% & 4.74\% \\ \midrule
$\max \lceil  W_\mathbf{H}(i)\rceil$ & 16 & 22  & 44  & 78  & 158  & 304  \\ \midrule
\# $\mathsf{XOR}$ gates              & 49 & 106 & 247 & 562 & 1247 & 2758 \\ \midrule
Execution ($\mathsf{pMult}()$) in cc & 5  & 6   & 7   & 8   & 9    & 10   \\ \bottomrule
\end{tabular}%
}
\end{table}

Finally, we give the proposed massive parallel GRAND variant for 5G NR control channels in Algorithm 2.

\begin{table}[h!]
	\centering
	\begin{tabular}{|c|}
		\hline
		\parbox{8.5cm}{\vspace{0.1cm}\textbf{Algorithm 2:} Massive Parallel GRAND for 5G \vspace{0.1cm}}\\
		\hline
		\parbox{8.5cm}{\vspace{0.1cm}{\bf Input:} Block length $N = 2^n$, received signal $Y^L = (r_1, \ldots, r_L)$ of $L$, $M$-QAM constellation points $r_i = (a_i, b_i)$ such that $L = \lceil \frac{N}{2m} \rceil$, where $M = 2^{2m}$; Hard-decoded demodulation sequence of $L$ constellation points $\hat{X}^L = (s_1, \ldots, s_L)$, where $s_i = (A_{ic}, A_{is})$; Parity-check matrix $\mathbf{H}_N$.\\
			{\bf Output:} Estimated codeword $\hat{\bm{x}}$ .
			\vspace{0.1cm}} \\
		\hline
		\parbox{8.5cm}{
			\begin{description}
                    \item[Step 0:] Initialize $\hat{L} \leftarrow \underbrace{(-, \ldots, -)}_L$
				\item[Step 1: For each]  $r_i, s_i, i \in 1, \ldots L$ \textbf{do}

                        \item \begin{description}
                            \item[ \ \ \ \ \ \ \ \ \ ] $\hat{L}_i \leftarrow (\mathcal{L}(r_i, s_i), s_i, s_{i d_1}, s_{i d_2}, s_{i {d_1}{d_2}}) $
                        \end{description}

                    \item[\ \ \ \ \ \ \ \ ] \textbf{End for}
                    \item[Step 2:] Sort $\hat{L} \leftarrow \mathsf{pSort}(\hat{L}, \mathsf{ascend}) $
                    \item[Step 3:] $\hat{L} \leftarrow  \{L_1, \ldots, L_S\} $, Cut-off the first $S$ constallation points 
                    \item[Step 4:] $\hat{Z} \leftarrow \text{All test error patterns from } \hat{L}$ 
                    
				\item[Step 5: While $syn \neq 0$ do] (Parallel codebook membership)
				
				        \item \begin{description}
					    \item[\ \ \ \ \ \ \ \ \ \ \ \ \ For each] $Z^L \in \hat{Z}$ \textbf{do}

                                \item \begin{description}
                                    \item[ \ \ \ \ \ \ \ \ \ ] $\hat{\bm{x}} \leftarrow \hat{X}^L \circledcirc Z^L$
                                    \item[ \ \ \ \ \ \ \ \ ] $syn \leftarrow \mathsf{pMult}(\mathbf{H}_N, \hat{\bm{x}})$
                                \end{description}
        
                        \item[\ \ \ \ \ \ \ ] \textbf{End for}
         
				        \end{description}
				
				\item[\ \ \ \ \ \ \ \ \ \ End while] 
				\item[Return] $\hat{\bm{x}}$
			\end{description}
		}\\
		\hline
	\end{tabular}
	\label{Alg:MassParalGRAND}
\end{table}

\begin{proposition}
\label{Prop:UpperParallelDecodingTime}
For a cut-off parameter $S$ and a parallel implementation of Algorithm 1 in $4^S$ instances, Algorithm 2 ends the search for the estimated codeword $\hat{\bm{x}}$ in no more than $2n + 2S + 4$ clock cycles.
\end{proposition}

A sketch of the proof for Proposition \ref{Prop:UpperParallelDecodingTime} is as follows: Step 1 can be executed in $L$ parallel instances, where each instance executes two multiplications, one addition and one square root operation. A parallel sort in Step 2 can be executed in $\mathcal{O}(\log_2 L)$ steps; and since $L<N$, it is upper bounded by $\mathcal{O}(\log_2 N)$. Step 4 deals with up to $4^S = 2^{2S}$ error patterns that must be distributed in the massive parallel ensemble of $4^S$ instances of parallel matrix-vector multiplication circuits. Distribution of all error patterns takes up to $2S$ parallel steps. In Step 5, the assignment part $\hat{\bm{x}} \leftarrow \hat{X}^L \circledcirc Z^L$ already happened within Step 4, so the time-consuming part is $syn \leftarrow \mathsf{pMult}(\mathbf{H}_N, \hat{\bm{x}})$ which according to Equation (\ref{Eq:pMultComplexity}) is $1 + \max_{i \in 1, \ldots K} \lceil \log_2( W_\mathbf{H}(i)) \rceil $ and for the instances of used Polar code matrices, is upper bounded by $\log_2 N = n$. Summing the steps from all parts of the algorithm and assuming that every elementary step can be executed in one clock cycle gives us the value of $2n + 2S + 4$ clock cycles.

At the end of this section, we discuss the plausibility of the assumption made in Proposition \ref{Prop:UpperParallelDecodingTime} where we consider the number of parallel instances for the matrix-vector multiplication to be up to $4^S$. For the concrete value $S = 8$, the number of considered instances is thus $2^{16} = 65536$. To understand if this is realistic, let us consider the case for $n = 7$. Based on Table \ref{Table:SummaryHN}, we need 984 $\mathsf{AND}$ gates, and 247 $\mathsf{XOR}$ gates, which means $\approx$64.5 million $\mathsf{AND}$ gates and $\approx$16.2 million $\mathsf{XOR}$ gates for $2^{16}$ instances. The modern ASIC chips could have those number of gates.


\section{Performance Evaluation}
\label{Sec:PerformanceEvaluation}

\begin{figure*}[!htbp]
  \centerline{\subfigure[QPSK or 4-QAM modulation]{\includegraphics[width=0.325\textwidth]{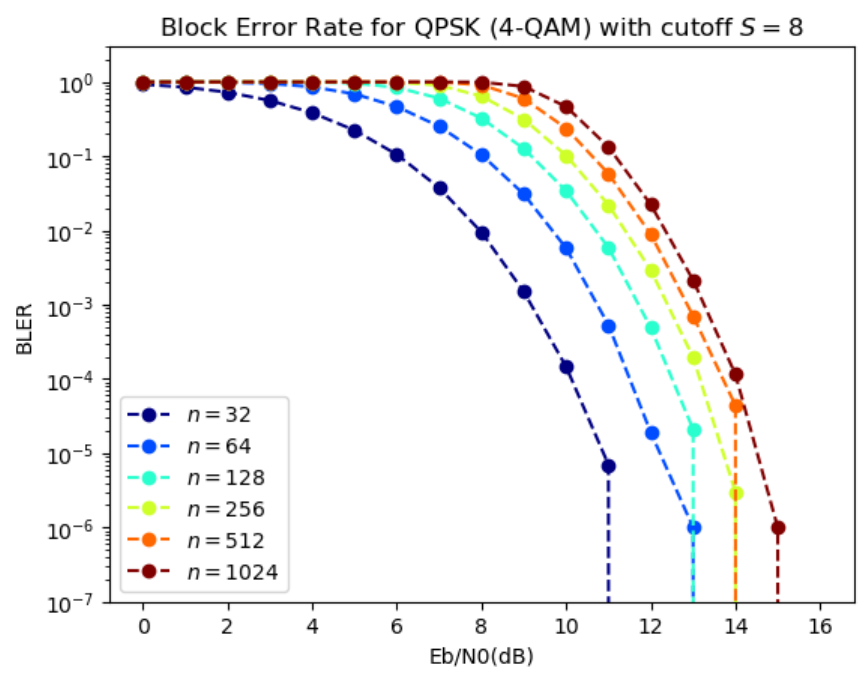}
      \label{4-QAM-Simulations}}
    \hfil
    \subfigure[16-QAM modulation]{\includegraphics[width=0.325\textwidth]{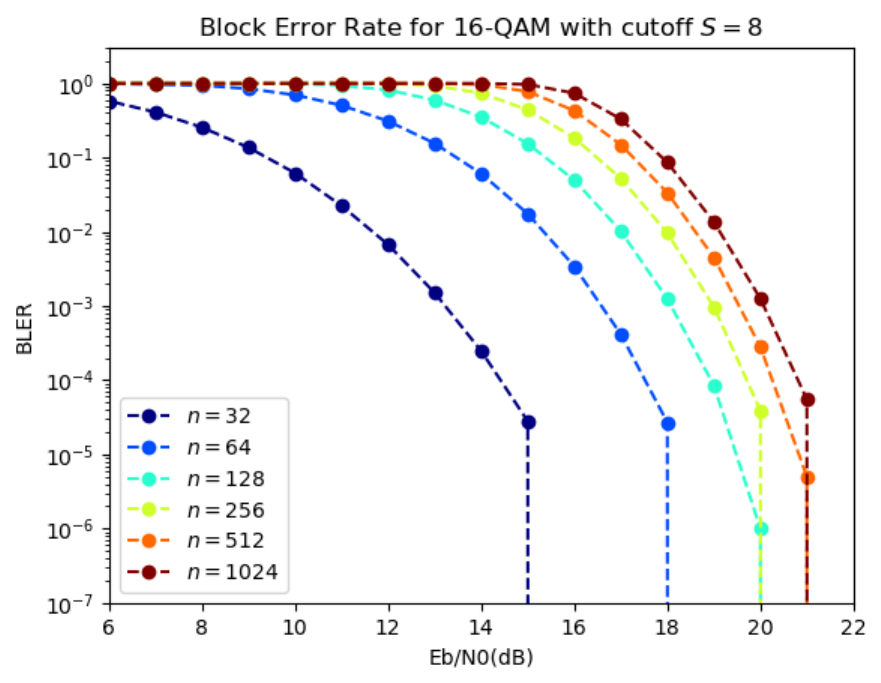}
      \label{16-QAM-Simulations}}
    \hfil 
    \subfigure[64-QAM modulation]{\includegraphics[width=0.325\textwidth]{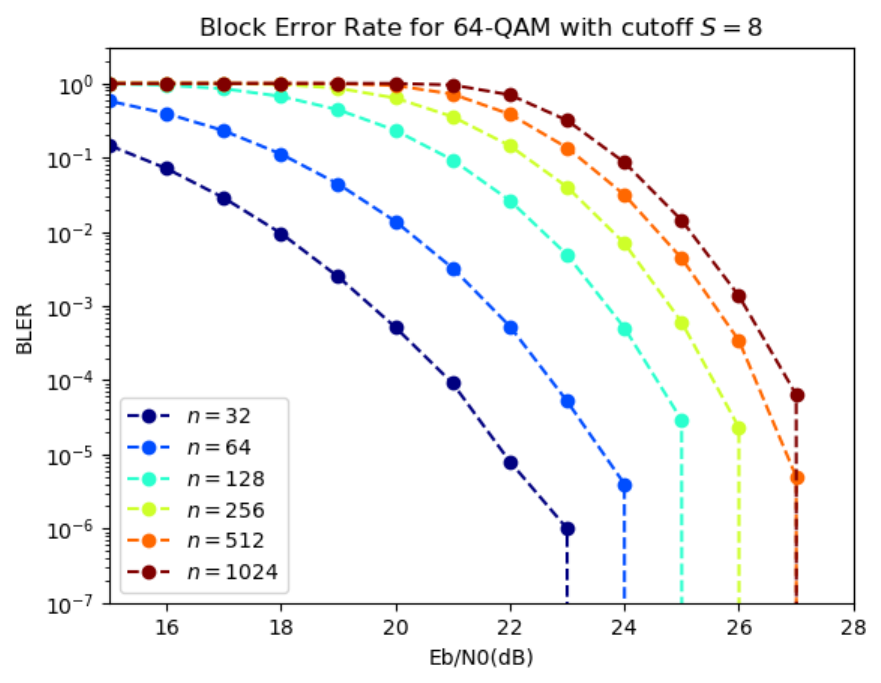}  
       \label{64-QAM-Simulations}}}
\end{figure*}

\begin{figure*}[!htbp]
  \centerline{\subfigure[$256$-QAM modulation]{\includegraphics[width=0.325\textwidth]{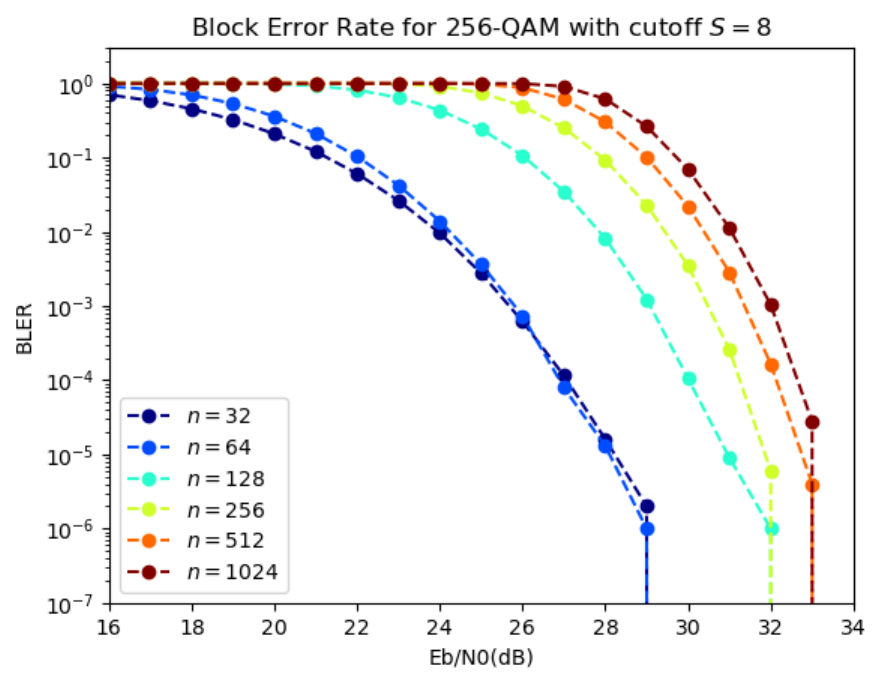}
      \label{256-QAM-Simulations}}
    \hfil
    \subfigure[$1024$-QAM modulation]{\includegraphics[width=0.325\textwidth]{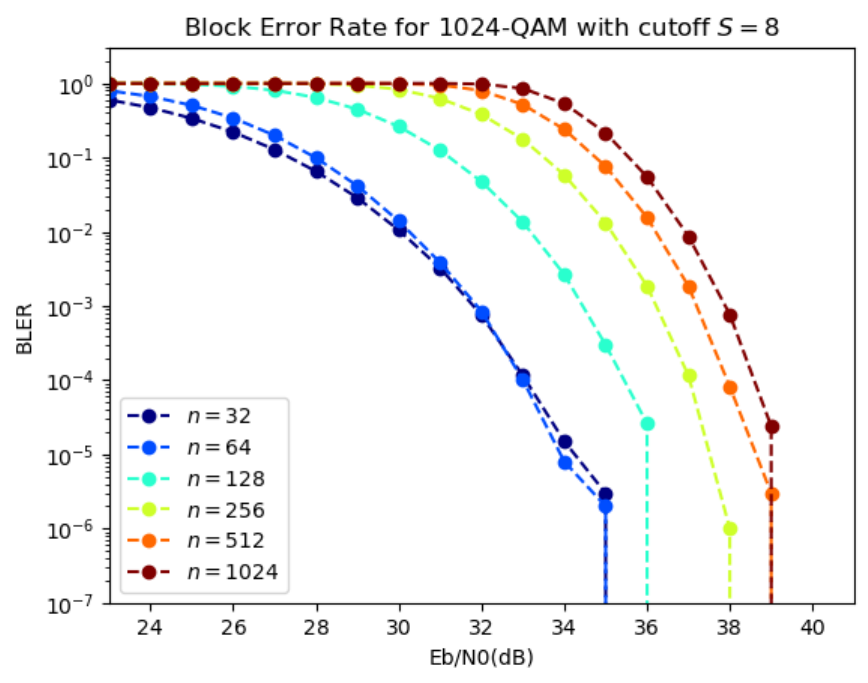}
      \label{1024-QAM-Simulations}}
    \hfil  
    \subfigure[$4096$-QAM modulation]{\includegraphics[width=0.325\textwidth]{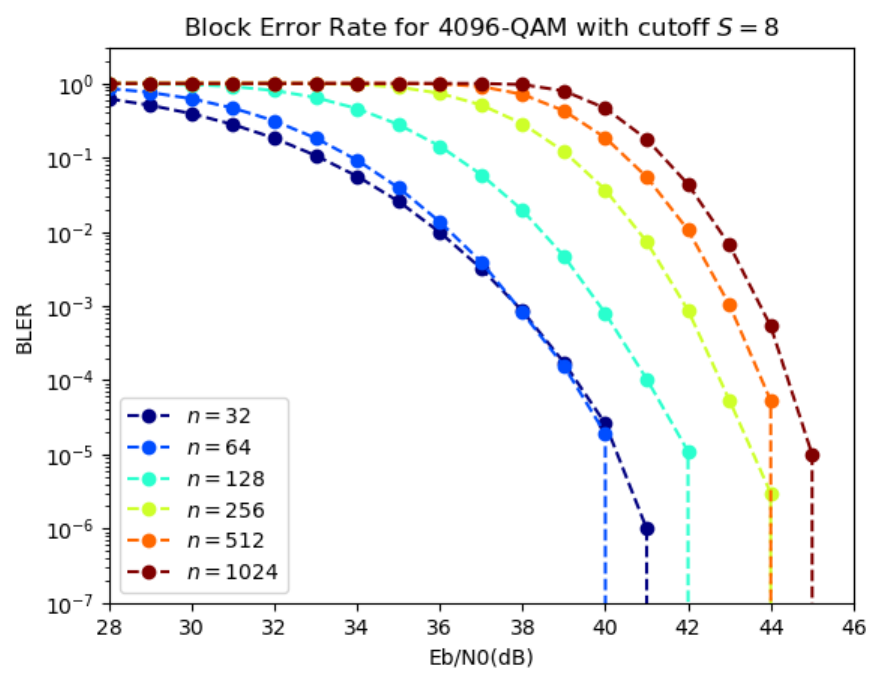}
      \label{4096-QAM-Simulations}}}
  \caption{Simulation results: Block Error Rate (BLER) for the proposed GRAND-like approach for decoding Polar codes with codeword lengths $N = 32, \ldots, 1024$ bits for different $M$-QAM modulations.}
  \label{fig:sim}
\end{figure*}

In this section, we present the empirical performance results of the proposed algorithm for higher-order $M$-QAM schemes with block length \(N \in \{32, \dots, 1024\}\) bits. The estimated Block Error Rate (BLER) points are computed using one million trials (for each plotted dot) for different $M$-QAM schemes and presented in Figure~\ref{fig:sim}. For the lower-order QAMs, a cutoff parameter value of $S = 8$ is optimal compared to higher-order QAMs since in the latter case, the number of received QAM signals for a codeword of given length $N = 2^n$ would be smaller than 8. For example, for $N = 32$ and 64-QAM, each signal represents 6 bits, and thus, the number of received signals to transmit a codeword of length 32 is 6 (the first integer value is 36 that is larger than $N$). For 256-QAM, since each signal carries 8 bits, to transmit a codeword of length 32 we need just 4 signals. As it is expected, the operating SNR regime increases for increasing orders $M$. 

Unlike the other GRAND approaches, our extensive experiments show that our approach is applicable for both short and long block lengths, up to 512 and 1024 bits, with great success. 

Whether short CA-Polar codes or longer codes, we completely agree with the positions expressed in the GRAND paper~\cite{an2022burst} that interleaving induces unwanted delays of the order of several thousands of bits. In our experiments, codewords were independently decoded from each other, exploiting the correlation in the noise and leveraging the modulation information from the high-order modulated symbols. 

In terms of achieving the goal of latency reduction, we differ from other GRAND approaches in the sense that similar approaches also try to achieve SNR gains, which are paid for by the increased complexity of decoding. However, in practice, the SNR operational ranges are predetermined for all modulation techniques, and our results are within those predetermined ranges. If reducing the latency is an important service level agreement (SLA) goal, it is highly unlikely that the network carriers would decrease their SNRs because of a potential 1-3 dB gain for some coding schemes.

\section{Conclusions}

In this paper, we proposed a massive parallel decoding framework using GRAND-like approach, focusing exclusively on extensive parallelization with the aim of achieving low latency in beyond 5G networks. All block lengths ($N$) specified for use in the 5G NR control channels were tested in this framework for higher-order modulation (M-QAM) techniques. We also introduced a novel likelihood function in the framework that effectively reduces the symbol error pattern space to $\mathcal{O}(4^{N/\log_2 M})$; thereby enabling the search for the estimated codeword using the massively parallel GRAND variant to be completed in at most $2\log_2 N + 2S + 4$ clock cycles. Results show that the proposed approach provides good BLER performance for the multiple M-QAM schemes and block lengths. 


\section*{Acknowledgment}
This work has received funding from the Research Council of Norway through the SFI Norwegian Centre for Cybersecurity in Critical Sectors (NORCICS) project \# 310105.

\nocite{*}

\enlargethispage{-1.2cm} 

\bibliographystyle{ieeetr}
\bibliography{bibliofile}

\end{document}